\documentclass[12pt,a4paper]{article}
\usepackage{amsfonts,amsmath,amsthm}
\usepackage{graphicx}
\usepackage{epsfig}
\usepackage{slashed}
\usepackage{amssymb}
\usepackage{color}

\newcommand{\be}{\begin{equation}}
\newcommand{\ee}{\end{equation}}
\newcommand{\bea}{\begin{eqnarray}}
\newcommand{\eea}{\end{eqnarray}}
\newcommand{\beq}{\begin{eqnarray}}
\newcommand{\eeq}{\end{eqnarray}}

\def\G{{\Gamma}}
\def\tk{\varkappa}

\def\sD{\slashed{D}}

\def\({\left(}
\def\){\right)}
\def\[{\left[}
\def\]{\right]}
\def\<{\left<}
\def\>{\right>}
\def\a{\alpha}

\def\g{\gamma}                  

\def\om{\omega}

\def\S{\sigma_{{22}}}

\def\tr{{\mathop{\rm tr}}}

\begin{document}

\title{
Charge density and conductivity of disordered Berry-Mondragon graphene nanoribbons\\
}

\author{{C. G. Beneventano$^\dag$, I. V. Fialkovsky$^\ddag$\footnote{ifialk@gmail.com}, E. M. Santangelo$^\dag$ and
D. V. Vassilevich$^\ddag$}\\{\it $^\dag$ Departamento de F\'isica, Universidad Nacional de La Plata}\\
{\it Instituto de F\'isica de La Plata,  CONICET--Universidad Nacional de La Plata,}\\
{\it C.C.67, 1900 La Plata, Argentina}\\
{\it $^\ddag$ CMCC-Universidade Federal do ABC, Santo Andr\'e, S.P., Brazil}
}

\maketitle

\begin{abstract}
We consider gated graphene nanoribbons subject to Berry-Mondragon boundary conditions in the presence of weak impurities. Using field--theoretical methods, we calculate the density of charge carriers (and, thus, the quantum capacitance) as well as the optical and DC conductivities at zero temperature. We discuss in detail their dependence on the gate (chemical) potential, and reveal a non-linear behaviour induced by the quantization of the transversal momentum. 

\end{abstract}




\section{Introduction}

It is not necessary nowadays to give any detailed presentation to graphene --- a monoatomic layer of $\pi$-bonded carbon atoms. It exhibits numerous exceptional properties, to which many a research is devoted. The interested reader is referred to the  abundant reviews on the subject. Some useful ones are \cite{grarev,RMP,RevDasSarma11}, and the recent book \cite{KatsBook}.

The fact that electron transport in graphene is described, in the continuum limit, by a massless Dirac equation was predicted theoretically \cite{Semenoff84,DiVincenzo84} more than twenty years before experimental confirmation \cite{1}. Since then, many of the predictions of a ``relativistic'' massless Dirac theory as applied to this material have indeed been confirmed.

When infinite, graphene behaves as a zero-gap semiconductor. Such behavior is a major obstacle for further application of this -- otherwise quite appealing -- two-dimensional material to electronic devices. This fact provoked a number of both theoretical and experimental studies aimed at engineering an energy gap in graphene.
Among possible mechanisms to achieve this goal, one notes strain--induced gap \cite{3} or induction by chemical modification of graphene \cite{4}. However, the use of samples of finite size would be the first natural try to actually induce a gap. Indeed, it was shown already in \cite{OrigZZ} that, at least in semiconducting armchair graphene nanoribbons, there is a band gap inversely proportional to the width of the ribbon $W$. A more general discussion was also presented in \cite{Akhmerov08}. Moreover, in \cite{6} it was shown, on the basis of a first-principle calculation, that both armchair and zigzag graphene nanoribbons do present a nonzero and direct band gap. However, both in \cite{Akhmerov08} and \cite{6}, it was stated that the value of the gap should be dependent on the crystallographical orientation of the edges of the ribbon, which is in contradiction to some of the experimental results \cite{7}.

A number of different mechanisms were suggested to explain this discrepancy \cite{Sols07,Ponomarenko08,Martin09,Adam08}, and only their combination seems to be favoured by recent experiments \cite{12,13}. However, ``as yet, a complete theoretical understanding remains elusive''\cite{RevDasSarma11}.  Nanoribbons with edge disorder \cite{Areshkin07,Mucciolo09}  and smoothly varying width \cite{Kats07} were also the subject of active theoretical research. The influence of the electron--electron interaction was investigated in \cite{Fernandez07}, while the effects of long--range bulk disorder and warping were studied in \cite{Klos09}. An overview of the state of the art in the field of nanoribbons, along with further literature, can be found in several recently published review articles on the transport and electronic properties of graphene nanostructures, including nanoribbons \cite{RMP,RevDasSarma11,RevDubois09, RevStampfer2011}
\footnote{We deliberately do not review the vast field of research of the transport properties in infinite graphene and graphene-based nanostructures, referring the interested reader to the reviews dedicated to this very important issue \cite{RevDasSarma11,RevDubois09,RevStampfer2011}, and many others.}.

In this paper, we use the methods of quantum field theory (QFT) as applied to the continuum model of graphene. These methods have already proved themselves to be quite useful in the description of many aspects of graphene physics
(see, e.g.  \cite{Maria,nos} and the reviews \cite{Voz1,QFTGra})
Among them, one should mention the absorption of light by graphene \cite{GSC,abs1},
the Faraday effect \cite{Faraday0,Faraday}, and the Casimir
effect \cite{Cas1,Cas2}.

In the context of rigorous mathematical approaches, required by QFT, it is well known that, in order to have a self-adjoint Hamiltonian in a bounded region, appropriate boundary conditions (BCs) must be supplied. The first conditions studied for a planar Dirac system (in the completely different framework of hadron physics and neutrino billiards) were the Berry-Mondragon ones, also known as  MIT bag or infinite--mass BCs \cite{BerryM,MITbag}. Later on, after the discovery of graphene, other conditions were also considered: zigzag and armchair ones \cite{OrigZZ}. A more mathematical treatment of the problem of the choice of BCs can be found, e.g. in \cite{Falko04,BenSant10}.

Our main goal is to calculate the induced density of charge carriers (and thus the quantum capacitance) and the optical conductivity of a disordered graphene nanoribbon with Berry--Mondragon boundary conditions imposed on its edges. We consider these BCs since they are rather universal in the description of confined fermions, being independent of the crystallographical orientation of the boundary. In \cite{Akhmerov08}, they were shown to model a staggered boundary potential as well. As shown in \cite{BenSant10}, among the local BCs imposed in a single Dirac valley which lead to a self-adjoint Hamiltonian, these are the only ones which open a neat gap in the energy spectrum. Some previous calculations concerning Berry-Mondragon graphene nanoribbons, based on other techniques, were presented, for instance, in Refs \cite{TitovPRL06,Kats13}.

This article is organized as follows: Section \ref{sec-probl} is devoted to setting up our notation and conventions. In Section \ref{sec-spectr} we study the spectrum and the complete set of normal modes of the boundary problem and in Sec. \ref{sec-prop} we use them to construct the propagator of the quasiparticles in graphene. This is the main ingredient in our calculations of the mean charge and optical conductivity, which appear in Sections \ref{sec-ch} and \ref{sec-opt} respectively. Finally, in Section \ref{sec-DC} we study the DC limit of the conductivity.

Throughout the paper we use the natural units $c=\hbar=1$, unless otherwise stated. All our calculations are performed for a single fermion species, i.e., for one valley and one spin value. To obtain results for a real graphene sample, one has to introduce the degeneracy factor, $N=4$.

\section{Graphene nanoribbons}\label{sec-probl}

Let us consider a graphene nanoribon of width $W$ placed along the $y\equiv x^2$ axis.
In the continuum description of the electronic states, which has been found to be quite accurate \cite{grarev}, the behaviour of the wave function is governed by the Dirac equation \cite{Semenoff84,DiVincenzo84}. For a single Dirac cone (valley) its covariant form can be written as
\begin{equation}\label{Dirac}
  \sD \psi(x)=0\,,
\end{equation}
where  $\psi=(\psi^1,\psi^2)$  is a two component spinor,
\begin{equation}
    \sD=i\tilde \gamma^\mu \partial_\mu\, +\,v_F m, \qquad
    (\tilde\gamma)\equiv (\gamma^0,v_F\gamma^1,v_F\gamma^2),
    \label{sD}
\end{equation}
and $\g^{0,1,2}$ are $2\times2$ gamma matrices in either of the two nonequivalent representations of the Clifford algebra in $2+1$ dimensions. Here and below we work in natural units, $\hbar=c=1$. In these units, the Fermi velocity $v_F\approx 1/300$.

The value of the mass gap parameter $m$ and the mechanisms of its generation
are under discussion \cite{Appelquist:1986fd,massgap1,Gusynin,Pyatkovskiy}.
For graphene on a substrate, the mass gap can reach $0.3$ eV \cite{Enderlein10,Zhou07}, while for suspended graphene samples it is usually very small. However, even in the latter case, the introduction of a mass parameter may be convenient on theoretical grounds, e.g., to perform the Pauli-Villars regularization (see Sections \ref{sec-ch} and \ref{sec-opt} in this paper).

As is well known, different representations can be selected for the gamma--matrices in (\ref{sD}) in a region without boundaries. Our boundary conditions will also be imposed in such a way that they preserve the invariance under unitary transformations between different representations. However, in order to perform the calculations in the forthcoming sections we have to be specific. Thus, we choose the following representation of the Dirac matrices in Minkowski space-time (with metric $(+,-,-)$)
\begin{equation}\label{gammas}
    \gamma^0=
    \left( \begin{array}{cc} 0 & 1 \\ 1 & 0 \end{array} \right),\quad
    \gamma^1=
    \left( \begin{array}{cc} -i & 0 \\ 0 & i \end{array} \right),\quad
    \gamma^2=
    \left( \begin{array}{cc} 0 & 1 \\ -1 & 0 \end{array} \right).
\end{equation}

The Berry--Modragon BCs were developed by considering fermions localized to a compact region due to a `locking' infinite mass potential \cite{BerryM}. They are the $2+1$ analogue of the so-called MIT bag BCs, introduced to model confinement in Quantum Chromodynamics in $3+1$ dimensions \cite{MITbag}, since they also imply zero current flux in the direction perpendicular to the boundary. Written in a $\gamma$-representation-independent way, they read
\begin{equation}
	\left.\frac{1+i{\gamma}^{\mu}n_{\mu}}{2}\psi \right|_{B}=0 \,,\nonumber
\end{equation}
where $B$ is the boundary of the region to be considered and $n_{\mu}$ is the $2+1$--dimensional external normal vector at the boundary.
In the case of an infinite ribbon along $x^2$, with our conventions, they read
\begin{eqnarray}
  \psi=-i\g^1 \psi  && {\rm at\ } x^1=0 \nonumber\\%
  \psi=+i\g^1 \psi  && {\rm at\ } x^1=W \label{BeM}.
\end{eqnarray}
In our particular representation (\ref{gammas}) these conditions can also be expressed in terms of the components of the bi-spinor $\psi=(\psi^1,\psi^2)$ as
\be
	\psi^1\big|_{x^1=0}=\psi^2\big|_{x^1=W}=0.
	\label{BeM2}
\ee


\section{Normal modes and spectrum of the boundary problem}%
\label{sec-spectr}

To deduce the spectrum and the normal modes of the above formulated boundary problem we consider an auxiliary selfadjoint operator $\mathcal{D}=\gamma^0\slashed{D}$, and solve its eigenvalue problem.

The eigenvalue equation $\mathcal{D}\psi = \mathcal{E}\psi$ is equivalent to
\begin{equation}
    \left( \begin{array}{cc} -i\partial_2 & -\partial_1 \\
        \partial_1 & i\partial_2
        \end{array}\right) \psi = \omega \psi-\gamma^0 m \psi \,,
    \label{eq-Om}
\end{equation}
with
\begin{equation}
\omega = (\mathcal{E}-k_0)/v_F,
\end{equation}
for $\psi\sim e^{-ik_0 x^0}$. The $x^0$-dependence will be omitted in what follows.

The $\psi^1$ component may be taken to satisfy explicitly the boundary condition (\ref{BeM2}) at $x^1=0$. Up to a normalization factor it reads
\begin{equation}
    \psi^1 =(\omega+k_2)\sin (k_1x^1) \exp (ik_2x^2).
    \label{psi_1}
\end{equation}
From (\ref{eq-Om}) we then find
\begin{equation}
    \psi^2 = \exp (ik_2x^2) \( k_1\cos (k_1x^1) + m \sin(k_1x^1)\)\,,
    \label{psi_2}
\end{equation}
with
\begin{equation}
    \omega = \a \varkappa, \qquad \varkappa=\sqrt{k_1^2+k_2^2+m^2},\qquad \a=\pm1 \,.
     \label{Omkk}
\end{equation}
The sign of $\a$ actually corresponds to electron/hole excitations in the ribbon.

Now, by using the explicit formulae for the spinor components (\ref{psi_1}), (\ref{psi_2}) we obtain, from the boundary condition at $x^1=W$, the following eigenvalue equation
\begin{equation}
    k_1\cos (k_1 W) + m \sin(k_1 W)=0.
    \label{eig}
\end{equation}
This equation can only be solved analytically in the zero mass case, giving
\be
    k_1=\frac{\pi}W \left( n-\frac 12 \right),\quad n=1,2,\dots
    \label{k_1_m}
\ee
Note that these allowed values of $k_1$ coincide with the ones determined, also in the massless case, through a study of the Hamiltonian in \cite{BenSant10,TitovPRL06}.
It is easy to see that, even for non-zero mass, the solutions $k_1(m)$ of (\ref{eig}) can be labelled by natural numbers and, thus, organized in an increasing order, $k_1^{(1)}(m)<k_1^{(2)}(m)<\ldots<k_1^{(n)}(m)<\ldots$

Using (\ref{psi_1}), (\ref{psi_2}) we choose normalized spinor eigenmodes in the following form
$$
    \psi_{\alpha,k_0,k_1,k_2} (x)= \frac{1}{2\pi}e^{-ik_0 x^0+ik_2x^2}\,\phi_{ \alpha,k_1,k_2} (x^1),
$$
\begin{equation}
\phi_{ \alpha,k_1,k_2}=
    \frac{1}{\(\varkappa(\varkappa+\a k_2)\(W+\frac{m}{m^2+k_1^2}\)\)^{1/2}}
        \( \begin{array}{c}
            \a(\varkappa+\a k_2) \sin (k_1x^1) \\
            k_1 \cos (k_1x^1)+m\sin(k_1x^1)
            \end{array}
        \)\,, \label{modes}
\end{equation}
where $k_1$ is solution of (\ref{eig}).
These modes obey the normalization condition
\begin{equation}
    \int d^3x\, \psi_{\alpha',k_0',k_1',k_2'}^\dag
    \psi_{\alpha,k_0,k_1,k_2}^{\phantom{\dag}}=\delta(k_0'-k_0)\delta(k'_2-k_2)
\delta_{k_1',k_1}\delta_{\alpha',\alpha}.
\end{equation}



\section{The fermion propagator}\label{sec-prop}

The QFT approach which we will employ in what follows is essentially based on the knowledge of the propagator of the fermionic quasiparticles in nanoribbons --- the inverse of (\ref{sD}), $\mathcal{S} \equiv \slashed{D}^{-1}$. In turn, $\slashed{D}^{-1}  = \mathcal{D}^{-1}\g_0$ and, thus, we can express $\mathcal{S}$ as a sum over the eigenmodes (\ref{modes}) of $ \mathcal{D}$
\begin{equation}
\mathcal{S}(x,y)\equiv \slashed{D}^{-1}(x,y)=
    \sum_{\alpha,k_1}\int dk_0\, dk_2\,
    \frac{\psi_{\alpha,k_0,k_1,k_2}(x)\otimes\bar\psi_{\alpha,k_0,k_1,k_2}(y)}
    {\mathcal{E}(\alpha,k_0,k_1,k_2)}.
    \label{S}
\end{equation}
The sum over $k_1$ goes along all of the solutions of the equation (\ref{eig}). The eigenvalues, $\mathcal{E}$, of $\mathcal{D}$, are
\begin{equation}
        \mathcal{E}(\alpha,k_0,k_1,k_2)=k_0 +\alpha v_F \varkappa\,.
    \label{E}
\end{equation}
Note that (\ref{S}) is a propagator of a single species of 2-spinors.

Next, to describe gated and disordered graphene nanoribbons we introduce a Fermi energy shift (or chemical potential), $\mu$, and a phenomenological parameter $\Gamma$ which effectively accounts for the presence of impurities. This can be done by means of the substitution \cite{Altland}
\begin{equation}\label{subs}
    k_0 \to \zeta(k_0)\equiv k_0 +\mu+ i \Gamma{\rm\ sgn}k_0, \quad \G>0,
\end{equation}
in $\mathcal{E}$ (\ref{E}) and, thus, in the denominator of (\ref{S}). In the limit $\Gamma\to 0$ one recovers the usual Feynman propagator. For $\Gamma\ne 0$ the propagator $\mathcal{S}$ is not an analytic function
of $k_0$ due to the presence of ${\rm sgn}\, k_0$.
More generally,  $\Gamma$ could depend on the frequency, on an external magnetic field, etc.
We shall restrict ourselves to a constant $\Gamma$ \cite{G_cnst_1}.

Describing the disorder in such a simplified manner, we assume that the long--range impurities present in the graphene nanoribbons are sufficiently weak. Otherwise the states near the Dirac points get localized, and deviations from the Dirac dispersion should be taken into account. The behaviour of graphene in the presence of strong long-range impurities is studied in detail in \cite{DisOrd}.


\section{Mean charge density}\label{sec-ch}

Having calculated the fermion propagator $\mathcal{S}$ (\ref{S}), we can express the density  of charge carriers, $n(x)$, as
\footnote{%
This expression comes from considering the mean density of fermionic current, $\langle j^\mu\rangle=\langle\bar\psi\g^\mu\psi\rangle$, which in quantum field theory is defined as the functional derivative of the so-called effective action in the presence of an external electromagnetic potential $A_\mu$
\begin{equation}
\< j^\mu (x)\>=-\left[ \frac \delta{\delta A_\mu(x) }\, S_{\rm eff}(A)\right]_{A=0}\,,\qquad
    S_{\rm eff}(A)\equiv-i\ln \det \Big(\tilde \gamma^\mu (i\partial_\mu-e A_\mu) +v_F m\Big)\,.\label{jSeff}
\end{equation}
Here, $det$ stays for a functional determinant of a differential operator. The effective action, $ S_{\rm eff}(A)$, is obtained by a functional integration over the fermions in the Dirac model. This action is a sum of one-loop Feynman diagrams with an arbitrary number of external photons. For details, see \cite{QFTGra}.}
\begin{equation}
    n(x)   = -i \, {\rm tr} (\gamma^0 \mathcal{S}(x,x))\,.\label{j01}
\end{equation}

We note that the substitution (\ref{subs}) with $\Gamma\ne 0$ actually breaks gauge invariance. To restore the invariance one has to add $eA_0$ to any fermion frequency, including the one which appears in ${\rm sgn}\, (k_0)$. This leads to new vertices
involving $A_0$ and proportional to $\Gamma$ \cite{BFV}. These new vertices contribute
to the mean charge density (but not to the conductivity). To avoid unnecessary complications, in this section we consider the case $\Gamma\to 0$ only.

By using (\ref{S}) we obtain from (\ref{j01})
\begin{equation}
n(x) =-i \sum_{\alpha,k_1}\int dk_0\, dk_2
\frac {\psi^\dag_{\alpha,k_0,k_1,k_2}(x)\psi_{\alpha,k_0,k_1,k_2}(x)}{\mathcal{E}(\alpha,k_0,k_1,k_2)}\,.
\label{j02}
\end{equation}

We shall be interested in the density of carriers averaged over the cross section of the nanoribbon,
\begin{equation}
    n\equiv \frac 1W\int_0^Wdx^1\, n(x) \,,\label{j03}
\end{equation}
which, with the help of the identity
\begin{equation}
    \int_0^Wdx^1\, \psi^\dag_{\alpha,k_0,k_1,k_2}(x)\psi_{\alpha,k_0,k_1,k_2}(x) =\frac 1{(2\pi)^2}\,,
\label{id}
\end{equation}
can be expressed as
\begin{equation}
   n=-\frac{i }{(2\pi)^2W} \sum_{\alpha,k_1}\int dk_0\, dk_2\,
        \frac 1{\mathcal{E}(\alpha,k_0,k_1,k_2)} \,.\label{j04}
\end{equation}

This expression is divergent and has to be renormalized. To this end we use the Pauli--Villars prescription, according to which one has to subtract from the integrand/summand in (\ref{j04}) the same expression, but with the mass $m$ replaced by a large mass parameter $M$. After calculating all integrals and sums, the limit $M\to\infty$ has to be taken. Clearly, if the resulting quantity is finite, it also vanishes at $m\to\infty$. This is the physical motivation for the Pauli--Villars subtraction scheme: for a very
large mass gap all fluctuations are frozen and do not contribute to quantum effects, as the mean charge density and conductivity, for example.
A similar idea was actually used by Berry and Mondragon \cite{BerryM} in deriving the `infinite mass' boundary conditions: to confine the fermions to a region their mass was considered to be infinite in the contradomain.

Therefore, the regularized charge density reads
\begin{equation}
    n_{\rm reg}=-\frac{i }{(2\pi)^2W} \sum_{\alpha,k_1}\int dk_0\, dk_2\,
    \left[\frac 1{\mathcal{E}(\alpha,k_0,k_1,k_2)} -\frac 1{\mathcal{E}(\alpha,k_0,k_1,k_2)_{m\to M}}\right]
 \,.\label{j05}
\end{equation}
here ${\mathcal{E}}=k_0+\mu+ i0^+{\rm sgn} k_0+\a v_F\tk(m)$.
The integral over $k_0$ can be easily performed by using the Cauchy theorem. The sum over $\alpha$
is also immediate. Assuming that $M$ is much larger than all
other dimensional parameters in the model, one obtains
\begin{equation}
    n_{\rm reg}=\frac{ {\rm sgn}\, \mu}{2\pi W} \sum_{k_1}\int_{-\infty}^{\infty}
     dk_2\,\Theta(|\mu|-v_F\tk(m))\,.
\label{j06}
\end{equation}
This expression does not depend on $M$, and the limit $M\to \infty$ is taken trivially. Therefore, eq.\ (\ref{j06})
in fact defines the renormalized density of carriers, $n^R$. After integrating over $k_2$, we have
\begin{equation}
    n^R=\frac{  {\rm sgn}\, \mu}{\pi W v_F} \sum_{k_1>0}
    \sqrt{\mu^2-v_F^2\tk_0^2}\,\Theta(\mu^2-v_F^2\tk_0^2)\,,
\label{j07}
\end{equation}
where $\tk_0\equiv\tk_0(m)=\sqrt{k_1^2(m)+m^2}$.

Note that, in the derivation above, the particular form of the dependence of $k_1=k_1^{(n)}(m)$ neither on index $n$ nor on the mass was essential. The same procedure is valid provided the $n$-th eigenvalue, $k_1^{(n)}(m)$, is a bounded function of $m$. Therefore, Eq.\ (\ref{j07}) is true for rather general boundary conditions, as long as there is no gapless mode.

Having derived the mean density of charge carriers, it is straightforward to calculate the quantum capacitance \cite{C_Qdef} as well. Using the definition, $C_Q=e \partial Q/\partial \mu$, where $Q$ is the induced charge density in the ribbon, we obtain, in full agreement with previous calculations \cite{Fang07,Shylau09},
\begin{equation}
    C_Q=\frac{e^2}{\pi W v_F} \sum_{k_1>0}
    	\frac{|\mu|}{\sqrt{\mu^2-v_F^2\tk_0^2}} \Theta(\mu^2-v_F^2\tk_0^2)\, .
\label{C_Q}
\end{equation}

The results (\ref{j07}), (\ref{C_Q}) coincide (up to the factor $N=4$, which accounts to spin and valley degeneracy in real graphene) with the ones obtained earlier via completely different approaches in \cite{Fang07,Shylau09}, where a detailed analysis of the capacitance and its dependence on the gate/chemical potential are also given.
Thus, the calculation presented above confirms once more the validity of the QFT approach to the description of transport properties in graphene systems.

\begin{figure}
\centering \includegraphics[width=8cm]{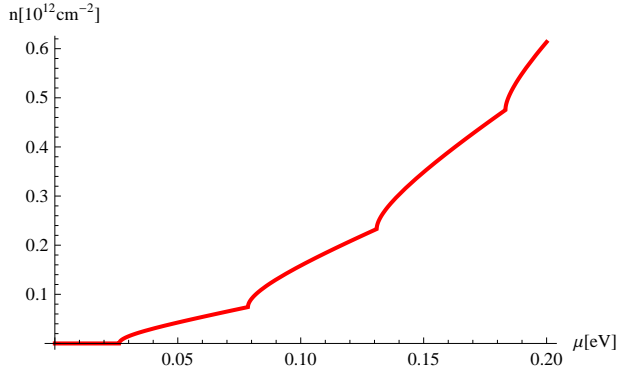}
\caption{
$n^R$ for a single fermion species as a function of the chemical potential $\mu$[eV], for $W=0.2$eV$^{-1}\simeq 40$nm, $m=0$.}
\label{on_mu}
\end{figure}

The behaviour of the charge carrier density as a function of the chemical potential is presented in Fig. \ref{on_mu}.


\section{Optical conductivity}\label{sec-opt}

To calculate the conductivity of graphene nanoribbons we shall need  \cite{QFTGra} the space--space components of the polarization tensor
\be
 \Pi^{jk}(x,y)    =
    i v_F^2 e^2  \tr\(\mathcal{S}(x,y) \gamma^j \mathcal{S}(y,x) \gamma^k\),
  \label{Pi_g}
\ee
where $j,k=1,2$, the propagator $\mathcal{S}$ is defined in (\ref{S}).
Once more, the calculations will be done for a single species of two-component fermions (i.e., for one spin and one valley). After performing a
Fourier transformation in the $x^0$ and $x^2$ directions we obtain
\be
\Pi^{jk}(\om,p_2;x^1,y^1)= \frac{ie^2 v_F^2}{(2 \pi)^2}\int dk_0dk_2
   \,\tr\[\mathcal{S}(k_0, k_2; x^1,y^1)\gamma^j \mathcal{S}(k_0-\om, k_2-p_2; y^1,x^1)\gamma^k\]\,,
    \label{Pi_g_Fou}
\ee
 and the Fourier representation of the propagator (\ref{S}) is
\begin{equation}\label{S_Four}
    \mathcal{S}(k_0, k_2; x^1,y^1)
     = 
     \sum_{\alpha k_1}
        \frac{\phi_{ \alpha,k_1,k_2}(x^1)\otimes\bar\phi_{\alpha,k_1,k_2}(y^1)}
        {\mathcal{E}(\alpha,k_0,k_1,k_2)}.
\end{equation}

Due to the lack of translation invariance along the $x^1$ direction we cannot further separate the $x^1$ dependence and, thus, the conductivity
will also depend on $x^1$. Since the measurements in nanoribbons do not yet allow to clearly resolve such a dependence, we shall consider the conductivity averaged across the width of the ribbon as done, for instance, in \cite{Kats13}.

The longitudinal component of the averaged conductivity tensor can be defined in terms of $\Pi^{jk}$ via the field-theoretical analogue of the Kubo formula as \cite{QFTGra}
\begin{equation}\label{s22_pre}
  \S(\om)=\frac{\Pi^{22}(\om)}{i\om}\,,
\end{equation}
where $\Pi^{22}(\om)$ is the corresponding component of (\ref{Pi_g_Fou}) taken at $p_2=0$ and averaged over the cross section of the ribbon
\be
    \Pi^{22}(\om)=
          \frac1W
            \int_0^W dx^1\int_0^W dy^1\Pi^{22}(\om,0;x^1,y^1).
   \label{s_pi}
\ee
Note that here one spatial integration corresponds to averaging along the cross--section of the ribbon, while the other one, effectively, to considering vanishing transversal momenta, $p_1=0$.

Using  (\ref{S_Four}) we can rewrite the above formula, identifying two (equivalent) interaction vertices
\be
\Pi^{22}(\om)=iC\int dk_0dk_2
    \sum_{\a k_1}\sum_{\a' k_1'}\frac1{{\cal E} {\cal E}'}
        \int_0^W dx^1 \bar\phi(x^1)\gamma^2\phi'(x^1)
        \int_0^W dy^1 \bar\phi'(y^1)\gamma^2\phi(y^1),
    \label{Pi22}
\ee
where $C=\frac{e^2 v_F^2}{(2 \pi)^2W}$, while $\phi(x^1)=\phi(\a,k_1, k_2;x^1)$ is given by (\ref{modes}), and
$$
\phi'\equiv \phi (\a',k_1',k_2;x^1),\qquad
    {\cal E}'\equiv  {\cal E}(\alpha',k_0-\om, k_1',k_2).
$$

The integration over $x^1$, $y^1$ in (\ref{Pi22}) is straightforward now. With the modes defined by (\ref{modes}) the vertices are simplified to
$$
\int_0^W dx^1 \bar\phi(x^1)\gamma^2\phi'(x^1)
    =-\frac{ \delta_{k_1 k_1'}}{ \varkappa}\( \a k_2\delta_{\a\a'}- \sqrt{k_1^2+m^2}\delta_{\a,-\a'} \).
$$
The right hand of this last equation is invariant under $\a \rightarrow \a'$; so, one gets the same result for both vertices in (\ref{Pi22}).

We arrive at the following expression for the longitudinal conductivity
\begin{eqnarray}
&&\S= \frac{C}{\om}\int dk_0dk_2
    \sum_{ k_1}\sum_{\a \a' } \frac1{\varkappa^2 {\cal E}  {\cal E} '}
        \( \a k_2\delta_{\a \a'}- \sqrt{k_1^2+m^2}\delta_{\a,- \a'} \)^2
\nonumber\\
&&\qquad
= \frac{C}{\om}\int dk_0dk_2
    \sum_{k_1}\sum_{\a\tilde\a }
        \frac{k_2^2 \delta_{\a\a'}+ \(k_1^2+m^2\)\delta_{\a,-\a'}}
                {\varkappa^2  {\cal E} {\cal E}'}
.\nonumber
\end{eqnarray}
After summation over  $\a$ and $\alpha'$, the above formula reduces to
\be
\S(\om)= \frac{2C}{\om}\int dk_0dk_2
    \sum_{k_1} \frac{\zeta(k_0)\zeta(k_0-\om)+v_F^2(k_2^2-k_1^2-m^2)}{\(\zeta^2(k_0)-v_F^2\varkappa^2\) \(\zeta^2(k_0-\om) -v_F^2\varkappa^2\)}\,,
    \label{S_xx}
\ee
where we used the fact that
${\cal E}_+{\cal E}_-{\cal E}_+'{\cal E}_-'=\(\zeta^2(k_0)-v_F^2\varkappa^2\) \(\zeta^2(k_0-\om)-v_F^2\varkappa^2\)$, with $\mathcal{E_\pm}\equiv \mathcal{E}(\alpha=\pm,\dots)$. We remind that
 $\zeta(k_0)=k_0 +\mu+ i \Gamma{\rm\ sgn}k_0$.

As mentioned in Section \ref{sec-prop}, for $\G>0$ the propagator $\cal S$ is not an analytic function of $k_0$ any more. However, after dividing the interval into three parts according to the zeros of the argument of the sign function, the integral over $k_0$ can be performed explicitly to get
\be\label{S_22}
\S(\om)= \frac{2C}{\om}\int dk_2\sum_{k_1}
        \(
            F +v_F^2(m^2+k_1^2)G
        \)\,,
\ee
where
\begin{eqnarray}
 &&  F =-\frac{i \G}{\om(\om+2i\G)}
        \log\frac{(v_F\tk+\mu)^2+\G^2}{(v_F\tk+\mu)^2-(\om+i\G)^2}
        +(\mu\to-\mu),
        \label{FG}\\
&&
G= \frac{4 i \pi}{v_F\tk\(\om^2-4v_F^2\tk^2\)}\nonumber\\
&&\qquad
    +\frac{1}{v_F\tk} \(
        \frac{\log\(v_F\tk-i\G+\mu\)-\log\({v_F\tk+\om+i\G+\mu}\) }{\om \(2v_F\tk+\om\)}\right.\nonumber\\
&&\qquad
            +\frac{\log\(v_F\tk+i\G+\mu\)-\log\({v_F\tk-\om-i\G+\mu}\)}{\om \(2v_F\tk-\om\)}
\nonumber\\
&&\qquad            \left.+\frac{\log\frac{v_F\tk-\om-i\G+\mu}{v_F\tk-i\G+\mu}}{(\om+2 i \G)(2v_F\tk-\om-2 i\G)}
            +\frac{\log\frac{v_F\tk+\om+i\G+\mu}{v_F\tk+i\G+\mu}}{(\om+2 i \G)(2v_F\tk+\om+2 i\G)}
            +(\mu\to-\mu)\).\nonumber
\end{eqnarray}

In general, in $2+1$ dimensions the polarization tensor (\ref{Pi_g}) is power-counting UV divergent. In $\S$ (\ref{S_22}), such divergence shows up in the asymptotic behaviour of the functions $F$ and $G$
\be
    F\simeq -\frac{2 i \G}{v_F^2 \tk^2}+O(\tk^{-4}), \qquad
    G\simeq -\frac{i \pi}{v_F^3 \tk^3}+\frac{4 i \G}{v_F^4 \tk^4}+O(\tk^{-5})\,,
    \label{assy}
\ee
which shows that the summation/integration in (\ref{S_22}) is indeed divergent. This divergency shall be handled via the same Pauli--Villars subtraction at infinite mass, which we have already used for the calculation of the density of charge carriers in Sec.\ \ref{sec-ch}.

In doing so, we consider a difference between two polarization operators (\ref{Pi_g}) taken at different masses $m$ and $M$, where the latter shall be taken to infinity after the loop momenta calculation. In  terms of the conductivity, this requires the consideration of the difference between $\S(m)$ and $\S(M)$. To calculate the $M\to\infty$ limit, we first separate the $m$  and $M$ dependent terms by adding and subtracting the asymptotics (\ref{assy}) from under the summation/integration
\be
    \Delta\S\equiv \S(m)- \S(M)
        =\S^{R}-\tilde\sigma_{22}^{R} + \frac{2C}{\om}  \int dk_2\sum_{ n}\(\frac{2 i \G}{v_F^2 \tilde\tk^2}-\frac{2 i \G}{v_F^2\tk^2}\)
        \label{DS}
        \ee
        $$
            +\frac{2C}{\om}\int dk_2\sum_{ n}
            \((M^2+k_1^2)\(\frac{i \pi}{v_F \tilde\tk^3}-\frac{4 i \G}{v_F^2 \tilde\tk^4}\)
            -(m^2+k_1^2)\(\frac{i \pi}{v_F \tk^3}-\frac{4 i \G}{v_F^2 \tk^4}\)
            \).
$$
Here,
\be
    \S^{R}= \frac{2C}{\om} \int dk_2\sum_{k_1}
        \(
            F +\frac{2 i \G}{v_F^2\tk^2}+(m^2+k_1^2)\(G+\frac{i \pi}{v_F \tk^3}-\frac{4 i \G}{v_F^2 \tk^4}\)
        \)\,,
        \label{S_R}
\ee
and with tilde we denoted quantities dependent on $M$ instead of $m$, as e.g., $\tilde\tk\equiv\tk|_{m\to M}$ $=\sqrt{k_1^2(M)+k_2^2+M^2}$, etc.

After the integration over $k_2$, all terms in (\ref{DS}), except for $\sigma_{22}^R - \tilde\sigma^R_{22}$, cancel against each other. One can also show that $\tilde\sigma^R_{22}\to 0$ as $M\to\infty$. Thus, the renormalized conductivity is given by $\S^R$ (\ref{S_R})
\be
    \lim_{M\to\infty}\Delta\S=\S^{R}.
\ee
We note that this is again true under the very mild assumptions on the dependence of $k_1$ on $m$ and $n$ that have been discussed at the end of Sec.\ \ref{sec-ch}.

\begin{figure}
\centering
\includegraphics[width=8cm]{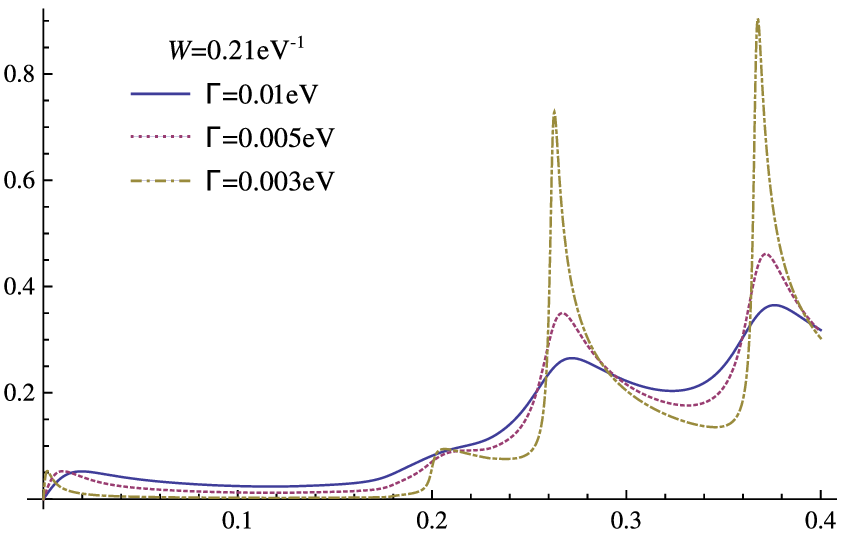}\includegraphics[width=8cm]{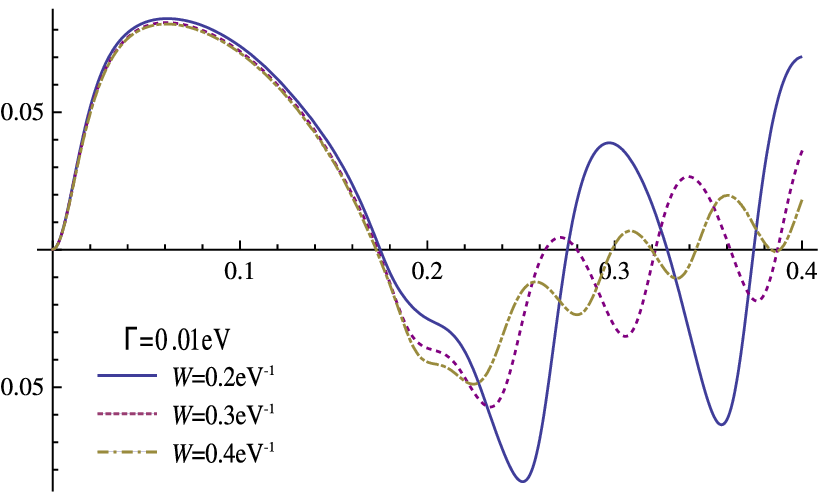}
\caption{
${\rm Re}\,\S^R $ (left) and ${\rm Im}\,\S^R $ (right) as functions of the frequency $\om$[eV] at fixed chemical potential $\mu=0.1$eV, for different values of $\G$ and $W$, and for $m=0$, in units of $2 e^2/h$.
}\label{ACcond}
\end{figure}

In figure \ref{ACcond} we present the real and imaginary parts of the optical conductivity (\ref{S_R}) as functions of the frequency at fixed chemical potential.
We can see that absorption in the nanoribbons considered here is comparatively small for frequencies smaller than $2\mu$, and then shows clear resonance lines. The peaks  in real and imaginary parts of conductivity correspond to those of the poles in the complex frequency plane lying at $\om = 2 v_F k_1^{(n)}-(1+2i)\G$.


\section{DC limit of the conductivity}\label{sec-DC}

To obtain the DC limit of the longitudinal conductivity, it is enough to expand the integrand of (\ref{S_R}) in a series for small frequencies, $\om\to0$
\be
    \S^{R}\mathop{\simeq}_{\om\to0} 4 \G^2 C \int dk_2\sum_{ n}
        \frac{v_F^4\tk^4 +2 v_F^2\tk^2(\G^2-\mu^2)+(\G^2+\mu^2)^2+8v_F^2\mu^2 k_2^2}
            {\(v_F^4\tk^4 +2 v_F^2\tk^2(\G^2-\mu^2)+(\G^2+\mu^2)^2\)^2 }+O(\om).
\ee
The integration over $k_2$ can be now performed explicitly, yielding
\be
    \S^{R}(\om=0)= \frac{ 2\pi C}{v_F \G\mu}\sum_{n}
        {\rm Im}\(\sqrt{v_F^2\tk_0^2(n)+(\G+i\mu)^2}- \frac{\G(\G+i\mu)}{\sqrt{v_F^2\tk_0^2(n)+(\G+i\mu)^2}}\)\,,
    \label{S_R0}
\ee
where $\tk_0 (n)=\sqrt{k_1^2(n)+m^2}$.

%
%

Let us consider the limit of small $\G$. The second term under the sum in (\ref{S_R0}) is $O(\sqrt{\G})$ at most, for all values of $\mu$,
while the first one is
\be
    {\rm Im}\sqrt{v_F^2\tk_0^2+(\G+i\mu)^2}\simeq {\rm sgn}\mu\,
    \Theta(\mu^2-v_F^2\tk_0^2 ) \sqrt{\mu^2-v_F^2\tk_0^2}
    +O(\sqrt{\G}).
    \label{LO}
\ee

Thus, one has the following expression for the conductivity
\be
    \S^{R}(\om=0)\simeq \frac{ 2\pi C}{v_F \G|\mu|}\sum_{k_1>0}
        \Theta(\mu^2-v_F^2\tk_0^2)\sqrt{\mu^2-v_F^2\tk_0^2}+O(1/\sqrt\G)\,.
    \label{S_app}
\ee
Note that, if $\mu^2$ is sufficiently far from the values $v_F^2(m^2 + k_1^2(n))$ for some $n$, the corrections in  (\ref{S_app}) are of the order $O(\G)$, rather than $O(1/\sqrt\G)$.
	
Thus, we see that the conductivity of an almost pure Berry-Mondragon graphene nanoribbon is quantized in a square-root manner. It shows an approximate gap, inversely proportional to the width of the ribbon, until $\mu$ reaches the first allowed value of the transversal momentum.

\begin{figure}
\centering \includegraphics[width=8cm]{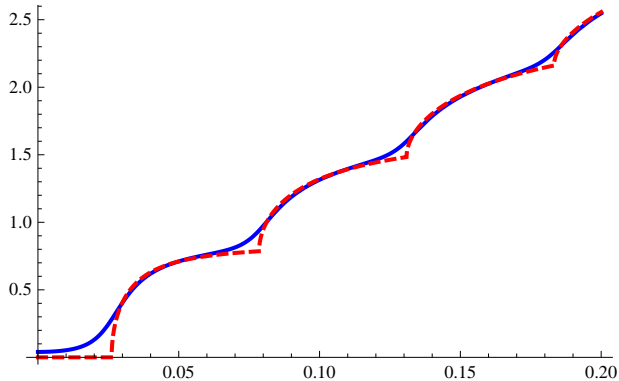}
\caption{$\S^R(0)$ for a single fermion specie as a function of chemical potential $\mu$[eV], for $W=0.2$eV$^{-1}\simeq 40$nm, $m=0$,  $\G=10$meV. Exact expression (\ref{S_R0}) in solid--blue, and approximate (\ref{S_app}) in dashed--red, in units of $2 e^2/h$.}
\label{DCcond}
\end{figure}

On Figure \ref{DCcond}, we show both the exact DC conductivity as given by equation (\ref{S_R0}), and its approximate value for small $\G$ (\ref{S_app}).

We note that the conductance of a real sample of graphene, $G=\S^{R} W /L$ (where $L$ is the  length of the sample), will depend on its width only through the values of the quantized transversal momenta $k_1$ (for $m=0$, these values are given by (\ref{k_1_m})), since $C\sim1/W$ in (\ref{S_R0}).

Having calculated the conductivity and the induced charge for almost clean Berry-Mondragon graphene nanorribbons, we can also deduce the mobility in this case,
\be
    \mathfrak{m}\equiv\frac{\S}{e n^R}=\frac{e v_F^2 }{2 \G \mu}\,,
\ee
which is independent of the width of the ribbon, and divergent in the limit of $\G\to0$.

As can be seen from Figure \ref{DCcond}, for vanishing values of the chemical potential (i.e. inside the gap), the conductivity reaches its minimal value, an analogue of the minimal conductivity in graphene. Indeed, taking in (\ref{S_R0}) the limit $\mu\to 0$ first, we obtain
\begin{equation}\label{Smin}
  \S^{min} = \frac{14C W^3 \G^2 \zeta(3)}{\pi^2 v_F^4} + O(\G^4)
   \simeq  0.13\,\frac{2e^2}{h}\,\frac{W^2 \G^2}{v_F^2}\,.
\end{equation}
In the last equality we restored the Plank constant $h$. Despite being proportional to $\G^2$, this minimal conductivity can reach relatively high values in the presence of disorder, due to the factor $v_F^{-2}$.
Note that the limit of an infinite two-dimensional sample, $W\to\infty$, is not appropriate after taking chemical potential to zero, $\mu\to0$ which, thus, is assumed to give the smallest scale in the problem. 

Finally, we  would like to emphasise that the obtained square--root dependence of the conductivity on the Fermi energy (chemical potential in our terminology) cannot be interpreted as smoothing of the unit--step one due to the presence of the impurities. It clearly follows from the approximate expression (\ref{S_app}) that in no sense our result approaches the unit--step function when $\G$ tends to zero.

\section{Conclusions and discussion}
In the present paper we have investigated gated graphene nanoribbons subject to Berry-Mondragon boundary conditions at zero temperature. We modelled the presence of disorder in the samples by introducing a phenomenological parameter --- the scattering rate $\G$.

Employing field--theoretical methods, as described in \cite{QFTGra} we calculated, in the first place, the induced mean density of charge carriers and the quantum capacitance in the limit of pristine graphene nanoribbons.  They are in accordance with earlier calculations \cite{Fang07,Shylau09}. In particular, van Hove singularities appear in the capacitance, positioned at those discrete values of the transversal momentum (times the Fermi velocity $v_F$) determined by the BCs. 
If measured, the separation between these singularities will allow for a direct observation of the band gap, $\Delta={v_F\pi}/{W}$.

In the second place, for disordered ribbons, we calculated both the optical and DC conductivities by the consideration of the polarization tensor. The optical conductivity shows an expected behaviour, with clear absorption lines at frequencies equal to double the discrete values of the transversal momentum (again, times the Fermi velocity $v_F$). However, in the DC limit, the conductivity 
shows distinct peculiarities as compared to the results of other authors obtained via Landauer approach.

On one hand, the transport gap obtained in our calculation is inversely proportional to the ribbon's width, $W$, similarly to previous theoretical \cite{OrigZZ,Akhmerov08,6} and experimental  results \cite{7,12}. But, instead of the step--like quantization of the conductance, obtained for the first time in \cite{Peres06}, see also \cite{Lian10}, we observe square--root steps (as a function of the chemical potential) with non--smoothness at those values corresponding to the quantized transversal momenta. Such strongly non--linear behaviour is most explicit for small values of the chemical potential, but smears out to a linear one in the asymptotic limit $\mu\to\infty$. Moreover, the steps we predict are not equally spaced (i.e., are not of equal height), contrary to what was found in the aforementioned articles.

The apparent discrepancy comes from the fact that  in \cite{Peres06} a clean system (both for zigzag and armchair BCs) was studied, and the transmission probability, $t$, which was found to give the units of the conductance quantization, was assumed `for simplicity' to be energy independent. However, this dependence appeared to be nontrivial, as was later revealed in \cite{TitovPRL06}, where the transmission probability (and thus the conductance) for ideal Berry--Mondragon and metallic armchair nanoribbons was calculated at non--zero gate potential. The latter paper mainly dealt with short ideal nanoribbons in the ballistic regime. This corresponds to the consideration of the $L\ll W, \G^{-1}$ limit ($L$ is the length of the ribbon). In this limit, the dependence of the transmission probability on the Fermi energy does indeed disappear (see Appendix A, \cite{TitovPRL06}).

On the other hand, the calculations presented here deal with the opposite limit, $L\gg \G^{-1}$, when the scattering length is assumed to be much shorter than the length of the ribbon. One can show that in the borderline case, i.e., when $L\sim \G^{-1}$, the conductance obtained using the results of \cite{TitovPRL06} (once averaged over the Fabri-Perot oscillations) behaves in accordance with the square--root quantization revealed in Section \ref{sec-DC} of the present paper. 

Experimentally, the conductance quantization (without magnetic field) in graphene nanoribbons was observed in \cite{Exp2,Lian10, Exp1}. While none of the experiments can resolve the form of the steps yet, all of them seem to favour equally spaced plateaux, a characteristic feature of the ballistic regime. However, the current precision of existing measurements cannot yet discriminate unambiguously between different predictions for quantization. Thus, we believe that further experimental investigation is crucial for gaining a complete understanding of the diffusive to ballistic transition in the transport properties of the graphene nano-ribbons.

Finally, we note that the obtained results for charge carriers density (\ref{j07}), quantum capacitance (\ref{C_Q}) and the conductivity (\ref{S_R}), (\ref{S_R0}) are expected to be valid for other BCs (which do not posses a gapless mode and do not mix transversal and longitudinal momenta) as well. 

\section*{Acknowledgements}
This work was supported in part by CNPq (D.V.V.) and FAPESP (I.V.F. and D.V.V.). Work of C.G.B. and E.M.S. was supported by UNLP (Proyecto 11/X615), CONICET (PIP1787) and ANPCyT (PICT909). The authors would like to thank M. Titov for fruitful discussions.


\end{document}